\documentstyle[12pt,psfig]{article}
\newcommand{\be}{\begin{equation}}
\newcommand{\ee}{\end{equation}}

\begin{document}
\normalsize
{\Large  Two Hydrodynamic Models of Granular Convection}
\vspace*{0.5cm}
\par\noindent
{\large Hisao Hayakawa }
\par\noindent
{\large  Graduate School of Human and Environmental Studies, 
Kyoto University, Kyoto 606-01 Japan} 
\par\vspace*{0.2cm}\noindent
{\large Daniel C. Hong }
\par\noindent
{\large Physics, Lewis Laboratory
Lehigh University, Bethlehem, PA 18015 USA}
\normalsize
\vspace*{2.5cm}

\bigskip
\noindent{\normalsize ABSTRACT:}
We present two continuum models A and B to study the convective instability of
granular materials subjected to vibrations.  We carry out the linear
stability analysis for model A and uncover the instability
mechanism as a supercritical bifurcation of a bouncing
solution.  We also explicitly determine the onset of convection
as a function of control parameters.  The simulations results are in
excellent agreement with the stability analysis.  Additional feature of
the model B is the inclusion of the relaxation term in the momentum equation,
which appears to be crucial in
capturing what is missing in model A, in particular, 
in reproducing
experimental convection patterns for large aspect ratio, both horizontally,
in which case convective rolls move toward the surface, and vertically
in which case convective rolls survive near the wall but
are suppressed in the bulk region.
\vspace*{1.0cm}

\noindent 1. INTRODUCTION

\vspace*{0.5cm}
It was Faraday who discovered the convective instability in a vibrated
granular bed in 1831.  Initially the flat surface of the granular pile
develops a heap upon vibrations, along the surface of which grains roll
down causing small or large scale avalanches.  Once formed,
such a heap is stable, because of the simultaneous formation of
permanent convective rolls inside the heap.  
Unlike Rayleigh-Bernard convection in fluids, however,
the origin of this instability
has remained relatively unexplored since its discovery, but recently
two simultaneous push from experimental side (
Clement et al, 1992; Pak and Behringer, 1994) with the use of MRI
or X-ray method (Knight et al, 1993) from large scale computer simulations
based on the distinct
element method 
(Taguchi, 1992; Gallas et al, 1992)
have aided our understanding through visualization.   However, the theoretical
efforts(Haff,1983;
Bourzutschky and Miller, 1995)
to uncover the basic mechanism of this
convective instability have not been remarkable, still largely focused on
producing convective patterns through computer models and simulations.  We have
recently undertaken steps, based on two continuum models, 
to remedy this situation, which appear to have captured the essence of
granular convection.  Considering a potentially important 
industrial application of
size segregation and a recent evidence (Knight et al, 1993) of the convection
connection in conjunction with the conventionally
held reorganization of grains, we consider
the search for the origin of granular convection quite important.
This is a brief summary of our effort along this
direction.  For details, see Hayakawa et al(1995) and  Yue(1995).

\vspace*{1.0cm}

\noindent 2. MODEL A

\vspace*{0.5cm}

We have studied two models. Both are based on
Navier-Stokes type continuum eqs, but ignore
temperature equation assuming the existence of 
a global temperature throughout the bed.
We first present the model A.

The staring point of model A
is the recognition that the
most fundamental aspect of the vibrating bed,
apart from the obvious fixed bed solution with no external driving, is
the existence of a uniform bouncing of a collection of particles, a solid
or a fluidized block with no internal degrees of freedom.  
In such a case, the bouncing solution,
represented by the motion of a ball on a vibrating platform, satisfies
$\ddot z = (-1 + \Gamma \sin{\it t})\theta (-1 + \Gamma \sin{\it t})$
where $z$ is the vertical coordinate of the granular block and
the $\theta(x)=1$ for $x>0$ and $\theta(x)=0$ for otherwise.
Next, in the presence of internal degrees of freedom 
such as rotation and/or translation, 
we define two coarse-grained dynamical
variables: the 
density $\rho({\bf r},t)$ and the velocity {\bf v}({\bf
r},t) of the granular system.  In the box fixed frame, eq. then modifies into:
\begin{eqnarray}
\label{(1a)}
 \partial_t\rho + \nabla\cdot(\rho {\bf v}) &=& 0 \\
 \partial_t{\bf v} + ({\bf v}\cdot\nabla){\bf v} &=& \hat z(\Gamma \sin{\it t}
-1-\lambda)
- {\frac{1}{\rho}}\nabla P \nonumber \\
& &+ {\frac{1}{R}}[\nabla^2{\bf v} + \chi\nabla(
\nabla\cdot{\bf v})] 
\label{(1b)}
\end{eqnarray}
where $\hat z$ is the unit vector in the vertical 
direction and $\lambda$ is the Lagrange multiplier.  
$\lambda=0$ for free motion and $\lambda=\Gamma \sin t-1$ for stationary state.
Note that the first term in the right hand side of
 (\ref{(1b)})
is due to the uniform bouncing and the
third term is the energy dissipation effectively represented by
the Reynolds number $R$ and the bulk viscosity $\chi$.  The pressure term
$P$ requires some discussion(Hayakawa et al, 1995) but the
Van der Waals model
$ P = T\rho/(1-b\rho)$ is a reasonable choice,
where $T $
represents the effective temperature which might be
a global variable and $b$  is a constant of order unity.

\vskip 0.5 true cm
To check the validity of our picture, 
we have solved (\ref{(1a)}) and (\ref{(1b)}) numerically 
in two dimension with
no slip boundary conditions at the side walls as well as at the top and the 
bottom plates. Note that the top plate suppresses complicated surface motion
of vibrating beds and allows us to use the simplified picture.
Since the granular fluid is confined in a box, we
do not introduce $\lambda$ explicitly in the simulations.  
The
absence of $\lambda$ and the presence of the top wall is expected to cause the 
appearance of the bouncing solution for $\Gamma \le 1$ 
in contrast with the real
situation but its omission would not change the essence of the
dynamics.  In the same spirit, we have ignored $\chi$ and $b$ in our 
simulations.

For $\Gamma<\Gamma_c$,
the bouncing solution is expected to appear inside the bed and  
the density and the velocity at
a given point oscillates with the same frequency of the vibration.(
Fig.2)
Upon increasing $\Gamma$ further to $\Gamma = 1.2$, which is
beyond the predicted $\Gamma_c = 1.12$ determined by $\sigma_M(\Gamma_c)=0$, 
we find that the bouncing solution has disappeared 
and the permanent convective rolls have developed inside the bulk (Fig.3).
The wavelength of the most unstable mode by the linear stability analysis 
is about $q_m\approx 0.4$, which is
not far from the actual wavelength of the convective rolls: $q=2\pi/\lambda = 
2\pi/L\approx 0.6$.  

\vskip 1.0 true cm
\vspace*{0.5cm}
\noindent 3. MODEL B
\vskip 0.5 true cm
We now introduce another model as  Model B. 
 Although the mass conservation remains same as in (\ref{(1a)}), 
the momentum conservation (\ref{(1b)}) now changes into
\begin{eqnarray} 
\label{(2a)}
\partial_tv_x +  ({\bf v}\cdot\nabla)v_x & =&
-(c_0^2/\rho)\partial_x\rho + \mu\nabla^2 v_x \\
\partial_tv_z + ({\bf v}\cdot\nabla) v_z
&=&(V(\rho,t)-v_z))/\tau  \nonumber \\
& &- (c_0^2/\rho)\partial_z\rho + \mu\nabla^2 v_z 
\label{(2b)}
\end{eqnarray}
where $c_0^2\simeq T$ is the sound speed.
The difference between model A and B is the presence of a relaxation
term in the
$z$ direction (\ref{(2b)}), which is represented by an
average function $V(\rho)$ with the relaxation time $\tau$. The origin of such
a term has been
discussed in Hong et al(1994)
in an attempt to introduce correlations
among grains or voids in the diffusing void model(DVM).  In the DVM,
the void speed is only a function of the local density, namely $v_z=V(\rho)$ + 
diffusion term. However, 
a void is a compressible hydrodynamic
object that changes and adjusts its shape to
conform to the surrounding, not instantaneously, but in
a given time.  So, it may be more appropriate to write down the time dependent
equation for the velocity in a manner given by (\ref{(2b)}) than
simply assuming a fixed value at a given local density.   The presence
of such relaxation process may be effectively
equivalent to assuming a drag force acting on a void.

\vskip 0.5 true cm

These coupled  equations (\ref{(2a)}) and (\ref{(2b)})
are also known as the traffic model or
two-phase model for fluidized beds that have been widely used for mixtures of
gas and granular particles. 
Functions in the model may be inferred
from the Enskoq equation; namely $-v_z/\tau$ is the drag term imposed
externally on the particle.  In the case of no interstitial fluid, its 
origin lies in the frictions of the front and the rear glass of the
container and from the wall.  Further,
the Enskoq pressure, $T\rho(1+f(\rho)\rho/2)$ with $f(\rho)$
the correlation function, produces an extra term $V(\rho)$ in addition to
the hard sphere pressure $T\rho$.  In this case, the coefficient of $V(\rho)$
is proportional to the gravity $g$.
The net effect is for the void (or particle)
to adjust its speed, $v_z$,
around the average value $V(\rho)$ in a given time $\tau$.
While deriving the exact form
for the function $V(\rho)$ is nontrivial, we know it must be a decreasing
function of density and have a cut off at the closed packed density $\rho_c$.
Hence, we have chosen a simple form:
$V(\rho) = V_o(\rho)(-1 + \Gamma \sin(\omega t))$
$ V_o(\rho) = (\rho_c-\rho)^{\beta}
\theta(\rho_c-\rho)$ with $\theta$ function and $\beta \le 1$.
We have investigated
the eqs.(\ref{(2a)}) and (\ref{(2b)}) numerically.
The mechanism for the
convection seems to be similar to model A, 
namely the superciritcal bifurcation
of a bouncing solution.  However, 
two notable differences emerge. 
First, when the aspect ratio increases vertically from, say one to two, 
the convection rolls that initially occupied the whole box move
toward the surface as shown in Fig.3
and the motion of the particles are fairly confined
near the surface. 
This is consistent with the experiments and MD simulations.

Second, when the aspect ratio increases horizontally, then convective rolls
inside the bulk are suppressed and they
appear only near the wall, which is not shown
here.
This again is consistent with the MD simulation
results of Taguch(1992). Hence, the role of drag appears to be important in
granular convection.

\vskip 1.0 true cm 
\noindent 4. DISCUSSION
\vskip 0.5 true cm
First, the bouncing solution as a basic state for granular convection seems
to have been confirmed in the simulations of both models A and B.  

Second, the role of boundary conditions.  We have employed no slip boundary 
conditions at the walls and the plate.  Further, we have put the rigid wall at
the top and thus suppressed the surface motion.  Granular materials have been
shown to exhibit very different motion near the wall in a zet flow experiment
under gravity(Caram and Hong, 1993) and there has been
some attempt to use
the negative or positive slip to control the convective patterns 
(Bourzutschky and Miller, 1995).  More detailed studies to derive reasonable
boundary conditions at the wall are required.

Third, the role of interstitial fluid.  Our model A assumes no
interstitial fluid such as air, 
and it predicts a series of rolls for the vertically large
aspect ratio.  Model B on the other hand
predicts that the convection is suppressed in the
bulk region but is confined near the surface, which is in accordance with
experiments and with the results of the two-fluid model with an interstitial
fluid.  Perhaps, the origin of drag, whether it is coming from friction at
the walls, or from the viscous effect of the interstitial fluid, may not be
relevant once it is present.  The suppression of the convection in the bulk
is due to the locking
mechanism of grains for near closed packed density, which was taken into
account in model B by a cut off  
in $V(\rho)$, namely $V(\rho)=0$ for $\rho\ge\rho_c$.  
We need more extensive 
studies of both models A and B to make quantitative comparison with
experiments.
\vskip 1.0 true cm

HH is, in part, supported by
Grant-in-Aid for Scientific Research from Japanese Ministry of Education,
Science and Culture (No. 08226206 and 08740308).

\vskip 1.0 true cm

\noindent REFERENCES
\vskip 0.5 true cm

\noindent Bourzutschky, M. \&  Miller,J. 1995. {\it Granular 
convection in a vibrating fluid.} Phys. Rev. Lett. 
\vskip 0.2 true cm
74: 2216-2219.
\vskip 0.2 true cm
\noindent Caram,H., \& Hong, D. 1992. {\it Diffusing void 
model for granular flow.} Mod. Phys. Lett. B. 
Vol.(6):761-771.
\vskip 0.2 true cm
\noindent Clement,E., Duran J., \& Rajchenbach,J. 1992. 
{\it Experimental study of heaping in a two dimen-
sional sand pile.} Phys. Rev. Lett. 69:1189-
1192. 
\vskip 0.2 true cm
%
\vskip 0.2 true cm
\noindent Gallas,J., Hermann,H.,\& Sokolowski,S. 1992. 
{\it Convection cells in vibrating granular media.} 
Phys. Rev. Lett. 69: 1371-1374.
\vskip 0.2 true cm
\noindent Haff,P. 1983. {\it Granular flow as a fluid mechanical 
phenomena.}  J. Fluid. Mech. 134: 401 
\vskip 0.2 true cm
\noindent Hayakawa,H,S. Yue,\& Hong,D. 1995. {\it Hydrodynamic
description of granular convection.} Phys. Rev. 
Lett. 75:2328-2331.
\vskip 0.2 true cm

\noindent Hong et al. 1994. {\it Granular relaxation under
tapping and the traffic equation.} Phys. Rev. E. 50:
4123-4135.

\vskip 0.2 true cm
\noindent Knight, J., Jaeger.H, \& Nagel,S. 1993. {\it Vibration
induced size separation in granular media: 
convection connection} Phys. Rev. Lett.
70: 3728-3731.

\vskip 0.2 true cm
\noindent Taguch,Y-h. 1992. {\it New origin of a convective 
motion.} Phys. Rev. Lett. 69: 1367-1369.
\vskip 0.2 true cm
\noindent  Yue,S 1995. {\it Nonlinear phenomena in materials 
failure and granular dynamics.} Ph.D thesis, 
Lehigh University.

\vskip 2.0 true cm
\noindent Figure Captions
\vskip 1.0 true cm
\noindent Figure 1:The effective growth rate $\sigma_{eff}(q)$ as a function of
the wave number q for $\Gamma =1.05$(diamond),for which
$\sigma_{eff}(q)<0$ for all values of q, while
for $\Gamma=1.2>\Gamma_c = 1.12$, 
$\sigma_{eff}(q)$ becomes positive for a band of q(square).  
$\Gamma_c$ is determined by the
condition that the maximum of $\sigma_{eff}(q)$ becomes zero
 at $\Gamma_c$.(cross)
The parameters used are: $T_e=R=10$ and $L=10$.(Figure is missing.  Please see
PRL 75,2328, 1995)
\vskip 0.3 true cm
\noindent Figure 2: For $\Gamma =1.2>\Gamma_c=1.12$,
the bouncing solution becomes unstable and
the permanent convective rolls appear inside the box.  The arrows are the
velocity vectors pointing upward.
\vskip 0.3 true cm
\noindent Figure 3:For $\Gamma =1.2>\Gamma_c$ and for large aspect ratio along
the vertical, the convective rolls move toward the surface.
\vskip 0.3 true cm
\noindent Figure 4:For $\Gamma =1.2>\Gamma_c$ and for large aspect ratio 
along the horizontal,
the convective rolls in the bulk are suppressed and survive only near the 
wall.

\newpage
\centerline{\hbox{
\psfig{figure=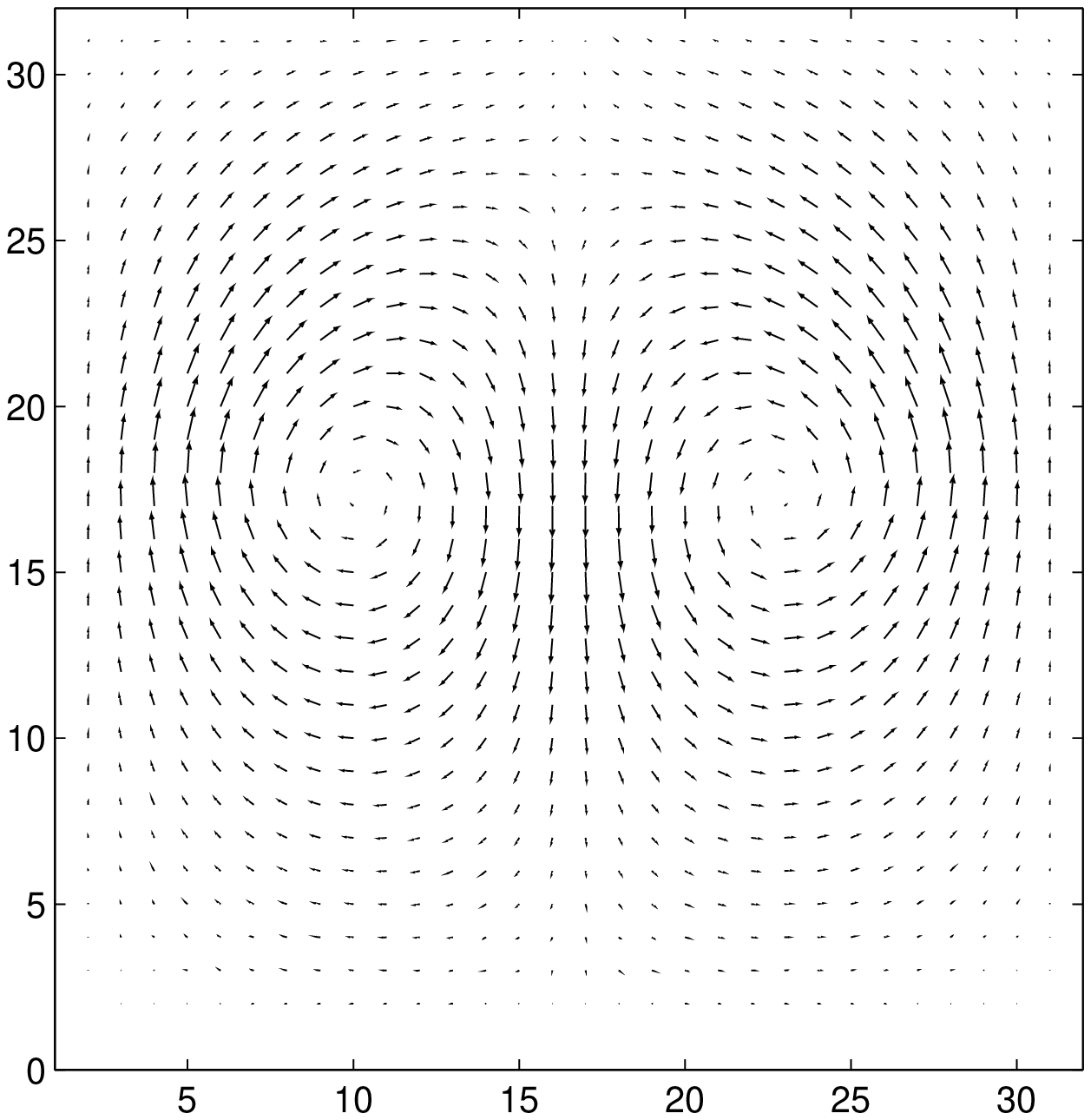}
}}

\newpage
\centerline{\hbox{
\psfig{figure=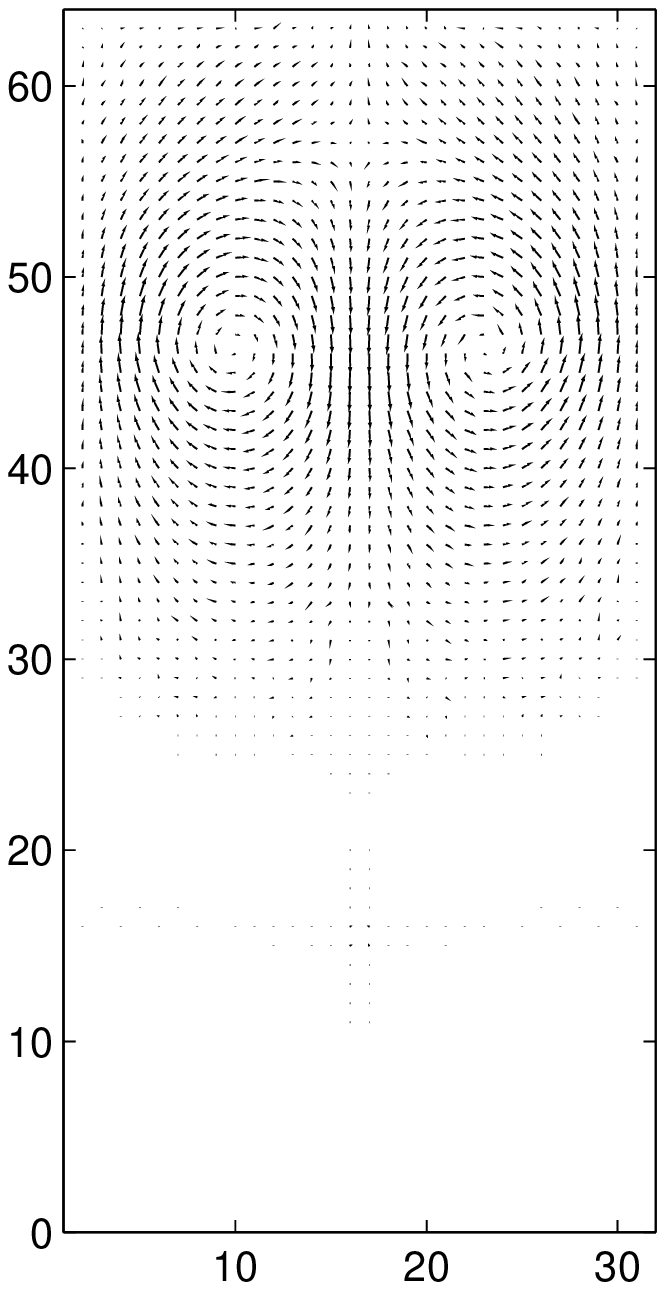}
}}

\newpage
\centerline{\hbox{
\psfig{figure=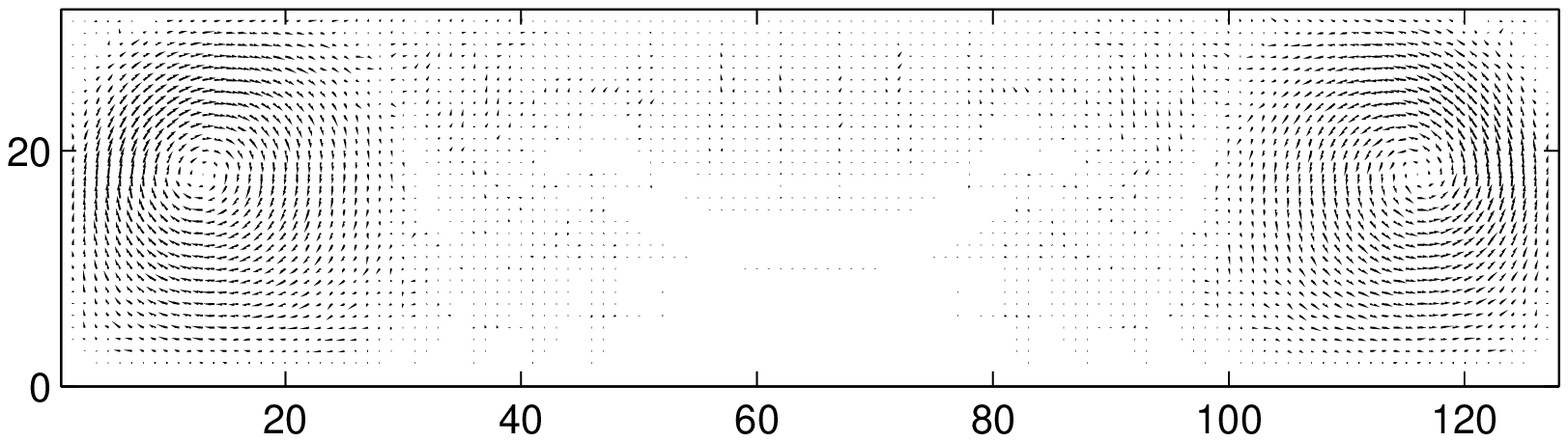}
}}

\end{document}